\documentclass[12pt]{article}
\usepackage[a4paper,top=1.5in,textwidth=16cm,textheight=22cm,left=3cm,twoside]{geometry}

\usepackage{fancyhdr}
\usepackage{copyright_instance}

\usepackage{hyperref} 
\usepackage{breakurl}
\usepackage{amssymb,amsmath}
\usepackage{mathrsfs}    
\usepackage{stmaryrd}    
\usepackage[ruled,vlined]{algorithm2e}
\usepackage{color}
\usepackage[all,dvips,arc,curve,color,frame]{xy}  
\CompileMatrices                                  
\usepackage{graphicx}
\usepackage{amssymb,amsmath,amsthm}
\SetKwInOut{Input}{input}
\SetKwInOut{Output}{output}
\SetKwInOut{DataStructures}{data structures}
\AlgoDontDisplayBlockMarkers\SetAlgoNoEnd\SetAlgoNoLine

\newtheorem{theorem}{Theorem}[section]

\newtheorem{corollary}[theorem]{Corollary}

\newtheorem{definition}[theorem]{Definition}
\SetKwProg{Fun}{function}{}{end}

\newtheorem*{pf}{Proof}

\newdimen\proofrulebreadth \proofrulebreadth=.05em
\newdimen\proofdotseparation \proofdotseparation=1.25ex
\newdimen\proofrulebaseline \proofrulebaseline=2ex
\newcount\proofdotnumber \proofdotnumber=3
\let\then\relax
\def\hfi{\hskip0pt plus.0001fil}
\mathchardef\squigto="3A3B
\newif\ifinsideprooftree\insideprooftreefalse
\newif\ifonleftofproofrule\onleftofproofrulefalse
\newif\ifproofdots\proofdotsfalse
\newif\ifdoubleproof\doubleprooffalse
\let\wereinproofbit\relax
\newdimen\shortenproofleft
\newdimen\shortenproofright
\newdimen\proofbelowshift
\newbox\proofabove
\newbox\proofbelow
\newbox\proofrulename
\def\shiftproofbelow{\let\next\relax\afterassignment\setshiftproofbelow\dimen0 }
\def\shiftproofbelowneg{\def\next{\multiply\dimen0 by-1 }%
\afterassignment\setshiftproofbelow\dimen0 }
\def\setshiftproofbelow{\next\proofbelowshift=\dimen0 }
\def\setproofrulebreadth{\proofrulebreadth}

\def\prooftree{
\ifnum  \lastpenalty=1
\then   \unpenalty
\else   \onleftofproofrulefalse
\fi
\ifonleftofproofrule
\else   \ifinsideprooftree
        \then   \hskip.5em plus1fil
        \fi
\fi
\bgroup
\setbox\proofbelow=\hbox{}\setbox\proofrulename=\hbox{}%
\let\justifies\proofover\let\leadsto\proofoverdots\let\Justifies\proofoverdbl
\let\using\proofusing\let\[\prooftree
\ifinsideprooftree\let\]\endprooftree\fi
\proofdotsfalse\doubleprooffalse
\let\thickness\setproofrulebreadth
\let\shiftright\shiftproofbelow \let\shift\shiftproofbelow
\let\shiftleft\shiftproofbelowneg
\let\ifwasinsideprooftree\ifinsideprooftree
\insideprooftreetrue
\setbox\proofabove=\hbox\bgroup$\displaystyle 
\let\wereinproofbit\prooftree
\shortenproofleft=0pt \shortenproofright=0pt \proofbelowshift=0pt
\onleftofproofruletrue\penalty1
}

\def\eproofbit{
\ifx    \wereinproofbit\prooftree
\then   \ifcase \lastpenalty
        \then   \shortenproofright=0pt  
        \or     \unpenalty\hfil         
        \or     \unpenalty\unskip       
        \else   \shortenproofright=0pt  
        \fi
\fi
\global\dimen0=\shortenproofleft
\global\dimen1=\shortenproofright
\global\dimen2=\proofrulebreadth
\global\dimen3=\proofbelowshift
\global\dimen4=\proofdotseparation
\global\count255=\proofdotnumber
$\egroup  
\shortenproofleft=\dimen0
\shortenproofright=\dimen1
\proofrulebreadth=\dimen2
\proofbelowshift=\dimen3
\proofdotseparation=\dimen4
\proofdotnumber=\count255
}

\def\proofover{
\eproofbit 
\setbox\proofbelow=\hbox\bgroup 
\let\wereinproofbit\proofover
$\displaystyle
}%
\def\proofoverdbl{
\eproofbit 
\doubleprooftrue
\setbox\proofbelow=\hbox\bgroup 
\let\wereinproofbit\proofoverdbl
$\displaystyle
}%
\def\proofoverdots{
\eproofbit 
\proofdotstrue
\setbox\proofbelow=\hbox\bgroup 
\let\wereinproofbit\proofoverdots
$\displaystyle
}%
\def\proofusing{
\eproofbit 
\setbox\proofrulename=\hbox\bgroup 
\let\wereinproofbit\proofusing
\kern0.3em$
}

\def\endprooftree{
\eproofbit 
  \dimen5 =0pt
\dimen0=\wd\proofabove \advance\dimen0-\shortenproofleft
\advance\dimen0-\shortenproofright
\dimen1=.5\dimen0 \advance\dimen1-.5\wd\proofbelow
\dimen4=\dimen1
\advance\dimen1\proofbelowshift \advance\dimen4-\proofbelowshift
\ifdim  \dimen1<0pt
\then   \advance\shortenproofleft\dimen1
        \advance\dimen0-\dimen1
        \dimen1=0pt
        \ifdim  \shortenproofleft<0pt
        \then   \setbox\proofabove=\hbox{%
                        \kern-\shortenproofleft\unhbox\proofabove}%
                \shortenproofleft=0pt
        \fi
\fi
\ifdim  \dimen4<0pt
\then   \advance\shortenproofright\dimen4
        \advance\dimen0-\dimen4
        \dimen4=0pt
\fi
\ifdim  \shortenproofright<\wd\proofrulename
\then   \shortenproofright=\wd\proofrulename
\fi
\dimen2=\shortenproofleft \advance\dimen2 by\dimen1
\dimen3=\shortenproofright\advance\dimen3 by\dimen4
\ifproofdots
\then
        \dimen6=\shortenproofleft \advance\dimen6 .5\dimen0
        \setbox1=\vbox to\proofdotseparation{\vss\hbox{$\cdot$}\vss}%
        \setbox0=\hbox{%
                \advance\dimen6-.5\wd1
                \kern\dimen6
                $\vcenter to\proofdotnumber\proofdotseparation
                        {\leaders\box1\vfill}$%
                \unhbox\proofrulename}%
\else   \dimen6=\fontdimen22\the\textfont2 
        \dimen7=\dimen6
        \advance\dimen6by.5\proofrulebreadth
        \advance\dimen7by-.5\proofrulebreadth
        \setbox0=\hbox{%
                \kern\shortenproofleft
                \ifdoubleproof
                \then   \hbox to\dimen0{%
                        $\mathsurround0pt\mathord=\mkern-6mu%
                        \cleaders\hbox{$\mkern-2mu=\mkern-2mu$}\hfill
                        \mkern-6mu\mathord=$}%
                \else   \vrule height\dimen6 depth-\dimen7 width\dimen0
                \fi
                \unhbox\proofrulename}%
        \ht0=\dimen6 \dp0=-\dimen7
\fi
\let\doll\relax
\ifwasinsideprooftree
\then   \let\VBOX\vbox
\else   \ifmmode\else$\let\doll=$\fi
        \let\VBOX\vcenter
\fi
\VBOX   {\baselineskip\proofrulebaseline \lineskip.2ex
        \expandafter\lineskiplimit\ifproofdots0ex\else-0.6ex\fi
        \hbox   spread\dimen5   {\hfi\unhbox\proofabove\hfi}%
        \hbox{\box0}%
        \hbox   {\kern\dimen2 \box\proofbelow}}\doll%
\global\dimen2=\dimen2
\global\dimen3=\dimen3
\egroup 
\ifonleftofproofrule
\then   \shortenproofleft=\dimen2
\fi
\shortenproofright=\dimen3
\onleftofproofrulefalse
\ifinsideprooftree
\then   \hskip.5em plus 1fil \penalty2
\fi
}

\definecolor{codeCol}{rgb}{0.4,0,0.4}

\newcommand{\pqProofForget}[1]{}
\newcommand{\pqTextForget}[1]{}
\newcommand{\pqLongText}[1]{}

\newcommand{\forget}[1]{}

\newcommand{\dotInItem}{\cdot}
\newcommand{\set}[1]{\mbox {$\{#1\}$}}

\newcommand{\pair}[2]{(#1 , #2 )}

\newcommand{\size}[1]{\mathop{\mid} #1 \mathop{\mid}}

\newcommand{\Fla}{\mathcal{L}\mathcal{A}}

\newcommand{\pqcomment}[1]{}

\newcommand{\SRauta}{SR-automata}
\newcommand{\SRauton}{SR-automaton}

\newcommand{\length}[1]{\mathop{\mid} #1 \mathop{\mid}}
\newcommand{\tauCA}{\tau_{c}}
\newcommand{\tauSR}{\tau_{sr}}
\newcommand{\Fopening}[2]{\mbox {${\rm{opening}}\pair{#1}{#2}$}}
\newcommand{\trConfig}[4]{\langle #1,\,#2,\,#3,\, #4 \rangle}
\newcommand{\trLabel}[3]{\squared{#1} : {#2} : {#3} }
\newcommand{\Fpush}[2]{\mbox {${\rm{push}}\pair{#1}{#2}$}}
\newcommand{\FpopN}[2]{\mbox {${\rm{pop}}^{#1}(#2)$}}
\newcommand{\Ftop}[1]{\mbox {${\rm{top}}({#1})$}}
\newcommand{\squared}[1]{[#1]}
\newcommand{\stacked}[1]{[{#1}]}
\newcommand{\Pinit}{P_I}
\newcommand{\Pfinal}{P_F}
\newcommand{\Ftree}[1]{\mbox {${\rm{tree}}({#1})$}}
\newcommand{\Flang}[1]{\mathcal{L}(#1)}

\renewcommand{\size}[1]{\mathop{\mid} #1 \mathop{\mid}}
\newcommand{\Qstates}{s}
\newcommand{\QstatesPrime}{s'}

\newcommand{\Qprod}{t}

\newcommand{\Nconfig}{\mathcal{C}}
\newcommand{\NconfigPrime}{\mathcal{C}'}
\newcommand{\NconfigDouble}{\mathcal{C}''}

\newcommand{\Fguess}[3]{\mbox {${\rm{guess}}({#1},{#2},{#3})$}}
\newcommand{\Fvalidate}[3]{\mbox {${\rm{validate}}({#1},{#2},{#3})$}}
\newcommand{\triple}[3]{(#1,#2,#3)}
\newcommand{\NsetOne}{GR}
\newcommand{\NsetTwo}{VR}
\newcommand{\NsetThree}{VQs}

\newcommand{\pqcut}[1]{}

\newcommand{\trArrow}{\rightsquigarrow}
\def\sArrow#1{\mathrel {\buildrel {#1}\over{\longrightarrow}}}
\def\sArrowSqStar#1%
  {\mathrel{{\buildrel{#1}\over{\longrightarrow}}^{\lower6pt\hbox{$\scriptstyle{*}$}}}}
\def\dArrow#1{\mathrel {\buildrel {#1}\over{\Longrightarrow}}}
\def\dArrowWord#1{\mathrel {\buildrel {#1}\over{\Longmapsto}}}

\title{
Walking on SR-automata to detect\\ grammar ambiguity
}

\author{Paola Quaglia\\ {\small{University of Trento}}}
\date{\mbox{}}

\begin{document}

\maketitle

\subsection*{Abstract}
We exploit the nondeterminism of LR parsing tables to reason about grammar
ambiguity after a conflict-driven strategy.
First, from parsing tables we define specialized structures, called SR-automata.
Next, we search for ambiguous words along the paths of SR-automata that reach a conflict
state and then diverge along the branches corresponding to distinct resolutions
of the conflict.

\

\noindent
{\fontsize{9.5}{12.5}\sffamily\bfseries {1998 ACM Subject Classification}}
{\fontsize{9.5}{12.5}\sffamily {F.4.2 Grammars and Other Rewriting Systems}}

\

\noindent
{\fontsize{9.5}{12.5}\sffamily\bfseries {Keywords}}
{\fontsize{9.5}{12.5}\sffamily {Context-free grammars; LR parsing; Ambiguity}}

\section{Introduction}

Grammar ambiguity is undecidable~\cite{Cantor62,Floyd62}, and
various, inevitably incomplete, approaches
have been investigated to detect ambiguity in some cases
(e.g., \cite{Gorn63,Amber01,Schmitz07,AxelssonHL08,BrabrandGM10,Schmitz10,Basten11}).
Some of these techniques are exploratory, meaning that ambiguous derivations
are searched for among those generated by the grammar.
Other methods are approximate, in the sense that the decision is taken
on some approximation of the given language.

Here we present a strategy for ambiguity detection that is centered around
a conflict-driven post-processing of the output of
a bottom-up parser.
We base our analysis on the widespread availability of LALR(1)~\cite{DeRemer69}
parser generators
(e.g.~\cite{YaccManual74,BisonManual}).
If the parsing table for a given grammar is deterministic, then
the grammar is surely unambiguous.
On the other hand, if the parsing table is nondeterministic,
then the grammar might be ambiguous, or it might belong
to a deterministic class bigger
than that for which the table was built.
Hence, we can let a parser generator do a pre-screening of unambiguity, and
perform further checks only on those grammars that lead to the construction
of nondeterministic tables.
Above we made the case for LALR(1) parsing tables.
The technique, however, applies to all the tables constructed as controllers
for the shift-reduce algorithm
(e.g., SLR(1)~\cite{DeRemer71}, LR(1)~\cite{AhoU77}).
So, in what follows, we generically refer to tables for LR parsing in its
broadest sense~\cite{Knuth65}, and indeed,
the bigger the class analyzable by the table,
the higher the probability that its nondeterminism depends on ambiguity.

To detect ambiguity, we focus on the conflicts found in nondeterministic tables.
First, we define \SRauta.
They are built from the characteristic automata underlying parsing tables,
and encode all the information needed to mimic the shift-reduce algorithm.
There is a main difference, though, between the two sorts of automata.
In the case of \SRauta, the accepted words are
obtained by a specialized concatenation of the terminals found along
an unbroken path from the initial to the final state.
This does not apply to characteristic automata, where, due to reductions,
the same words are recognized by concatenating the terminals scattered along
segmented paths.

Working on \SRauta, we look for ambiguous words among those that can be
recognized along paths that traverse a conflict state.
This activity is partially abstracted by operating on approximated versions
of \SRauta\ that are forgetful of the details needed to control executions,
and hence accept a superset of the language under investigation.
Essentially, we guess the ambiguous words by searching paths on the approximated
structure.
We then go back to the \SRauton\ to validate those words against
proper executions of the shift-reduce algorithm.

The rest of the paper is organized as follows.
Sec.~\ref{sec:preliminaries} presents basic definitions and conventions.
\SRauta\ and their properties are dealt with in Sec.~\ref{sec:srautomata}.
The proposed detection strategy is the subject of Sec.~\ref{sec:detection},
and Sec.~\ref{sec:conclusions} concludes the paper.
We assume the reader be familiar with the theory of LR parsing
(see, e.g.,~\cite{AhoLSU06,SippuS90}).

\section{Preliminaries}
\label{sec:preliminaries}

In this section, we will collect basic definitions and the adopted conventions.

A context-free grammar is a tuple $\mathcal{G} = (V,T,S,\mathcal{P})$ where
$V$ is the finite set of terminals and nonterminals,
$T$ is the set of terminals,
$S\in (V\setminus{T})$ is the start symbol, and
$\mathcal{P}$ is the finite set of productions.
We assume grammars be reduced, and adopt the following notational conventions.
The empty string is denoted by $\epsilon$,
$V^*$ is ranged over by $\alpha,\beta,\ldots$,
$(V \setminus T)$ by $A,B,\ldots$,
$T$ by $a,b,\ldots$,
$T\cup\set{\$}$ by $x,x',\ldots$, and
$T^*$ by $w,w',\ldots$.
Productions are written
$A\rightarrow\beta$, and
$\length{\beta}$ denotes the length of $\beta$.
Moreover, $\Flang{\mathcal{G}}$ stands for the language generated by
$\mathcal{G}$.

Given any context-free grammar $\mathcal{G}$,
LR parsing is applied to strings followed by the endmarker symbol $\$\notin V$.
The parsing table is constructed for the augmented version of $\mathcal{G}$
defined as
$(V',T',S',\mathcal{P}')$ where
$S'$ is a fresh symbol,
$V' = V\cup\set{S'}$,
$T' = T\cup\set{\$}$,
and $\mathcal{P}' = \mathcal{P}\cup\set{S'\rightarrow{S}\$}$.
Parsing is performed by running the shift-reduce algorithm~\cite{Knuth65}
using a parsing table as controller, and reading the next input symbol.
Two auxiliary structures are involved: a stack to trace the history of
computation by recording the traversed states, and a stack to keep trace
of the reductions performed.
When the parsing of a given word $w$ is successful, the second stack, named
\Ftree{w}, contains, from top to bottom, the sequence of productions for the
rightmost derivation of $w$ in $\mathcal{G}$.

Different controllers are adopted for different classes of LR parsing.
Nonetheless, in any case the parsing table is mechanically computed from two
objects that are finer or coarser depending on which class of grammars
the table is supposed to parse~\cite{bobu}.
These objects are
a \emph{characteristic automaton}, and
a \emph{lookahead function}.
Characteristic automata are deterministic finite state automata.
Their states are sets of items, i.e. of productions with a dot at some position
of their right-hand side.
The initial state is the one containing the item $S'\rightarrow\dotInItem{S}\$$,
and the final state is the one containing $S'\rightarrow{S}\$\dotInItem$.
The transition function of the characteristic automaton is used to set up
the \emph{shift} and the \emph{goto} entries of the parsing table.
A directive to \emph{reduce} the production $A\rightarrow{\beta}$
is inserted in the table at the entry $\pair{P}{x}$ iff
the state $P$ of the automaton contains the item $A\rightarrow{\beta}\dotInItem$
and $x\in\Fla\pair{A\rightarrow\beta}{P}$, where
$\Fla\pair{\_}{\_}$ is the lookahead function mentioned above.

Parsing tables can be nondeterministic, meaning that they may have multiply-defined
entries containing either a shift and a reduce directive (called s/r conflict) or
multiple reduce directives relative to distinct productions (called r/r conflict).
Here we are mainly interested in nondeterministic parsing tables.
To run the shift-reduce algorithm over them, we assume that,
any time control goes to an entry of the table that contains a conflict, a random local
choice resolves the conflict in favour of one of the possible alternatives.

In what follows,
given a characteristic automaton ${\mathcal{A}}$ for the augmented version
of a grammar $\mathcal{G}$
and an associated lookahead function $\Fla$,
we will refer to the pair $\pair{\mathcal{A}}{\Fla}$
as to a \emph{parsing table of} $\mathcal{G}$.
Also, we will call \emph{parsing of} $w$ \emph{on} $\pair{\mathcal{A}}{\Fla}$
the application to the string $w$ of the shift-reduce algorithm driven by
the controller $\pair{\mathcal{A}}{\Fla}$.

\section{\SRauta}
\label{sec:srautomata}

In this section, we will define \SRauta, and present their main properties.

Below, we will denote automata with a single final state by a tuple
whose elements represent, respectively,
the set of states,
the vocabulary,
the transition function,
the initial state,
and the final state.
Also, given a set $L$ of symbols, we let $\squared{L}$ represent the set of all
the elements of $L$ surrounded by square brackets.
We call \emph{prospective symbols} the elements of $\squared{L}$.
The intuition behind a prospective symbol like $\squared{x}$ is that
we go across it pretending
that its concrete counterpart $x$ will eventually be found and consumed.

\begin{definition}[\SRauta: layout]
  Let $\pair{\mathcal{A}}{\Fla}$ be a parsing table of
  $\mathcal{G} = (V,T,S,\mathcal{P})$.
  Also, let ${\mathcal{A}}=(\mathcal{Q},V\cup\set{\$},\tauCA,\Pinit,\Pfinal)$.
  Then the
  \emph{\SRauton\ for} $\pair{\mathcal{A}}{\Fla}$
  is the finite state automaton
  $(\mathcal{Q},
    T \cup \set{\$} \mathrel{\cup}
    (\squared{T\cup\set{\$}} \times {\mathcal{Q}} \times {\mathcal{P}}),
    \tauSR,\Pinit,\Pfinal)$
  where $\tauSR$ is defined by the following rules
  \begin{center}
  $\begin{array}{l@{\hspace{7ex}}l}
    \prooftree
        \tauCA\pair{P}{x} = Q
    \justifies
        \tauSR\pair{P}{x} = Q
    \endprooftree
  &  \prooftree
        x\in\Fla\pair{A\rightarrow\beta}{P} ,\quad
        R \in \Fopening{A\rightarrow\beta}{P} ,\quad
        \tauCA\pair{R}{A} = Q
    \justifies
        \tauSR\pair{P}{\trLabel{x}{R}{A\rightarrow\beta}} = Q
    \endprooftree
    \\[2ex]
  \end{array}$
  \end{center}
  with $R \in \Fopening{A\rightarrow\beta}{P}$ iff
  there is a path spelling $\beta$ in $\mathcal{A}$ from $R$ to $P$.
\end{definition}

\begin{figure}
  \centering
  \includegraphics[width=16cm]{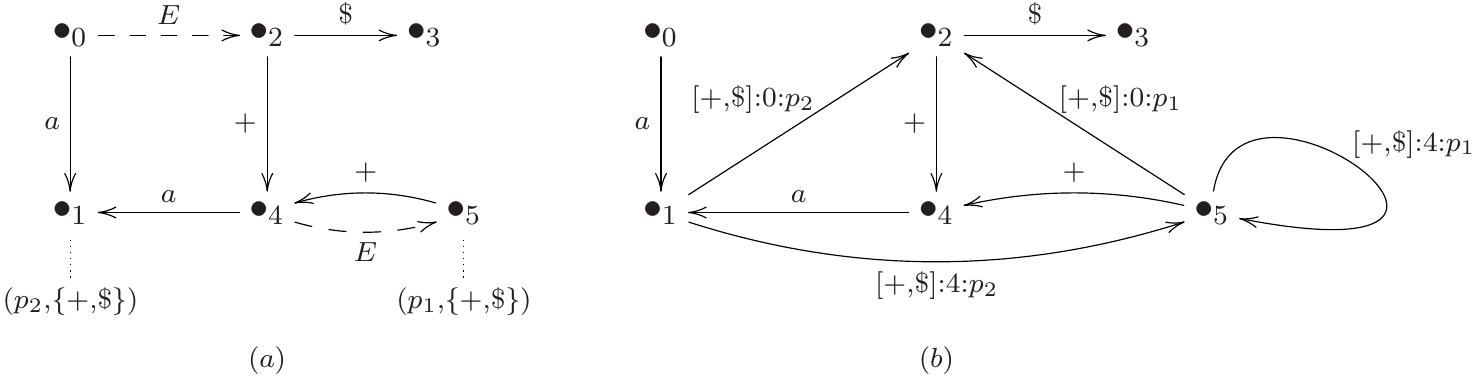}
\caption{\label{fig:G1}
          Parsing table of ${\mathcal{G}_1}$ \`{a} la Bison (a),
          and corresponding \SRauton\ (b).
          In both structures
          $\Pinit$ is state $0$,
          $p_1$ stands for ${E\rightarrow{E+E}}$, and
          $p_2$ for ${E\rightarrow{a}}$.
          In (a), the dotted line from $1$ to $(p_2,\set{+,\$})$
          means that in state $1$ the lookaheads $+$ and $\$$
          call for a reduction by $p_2$.
          The meaning of $(p_1,\set{+,\$})$ is analogous.
          In (b), labels like $[x_1,x_2]:n:p$ are shorthands for two edges,
          labelled by $[x_1]:n:p$ and by $[x_2]:n:p$, resp..
          }
\end{figure}

For the grammar $\mathcal{G}_1$ with production-set given by
$\set{E\rightarrow{E+E},\, E\rightarrow{a}}$,
Fig.~\ref{fig:G1} shows the instance of $\pair{\mathcal{A}}{\Fla}$
\`{a} la Bison and the corresponding \SRauton.

The language accepted by an \SRauton\ is defined in terms of an
\emph{execution relation}
which describes transitions between configurations.
Each configuration is a quadruple of the shape
$\trConfig{h}{\Qstates}{z}{\Qprod}$ where
$z$ is a string of symbols in $T\cup\set{\$}$,
$\Qstates$ (called state-stack) is a stack that contains states,
$\Qprod$  (called production-stack) is a stack that contains productions,
and $h$ is an auxiliary object to trace which prospective symbol, if any,
was involved in the past transition.
We denote stacks as lists of elements with the top of stack at the rightmost position,
so that $\squared{}$ represents the empty stack.
Also, we use the usual functions \Ftop{\_}, \Fpush{\_}{\_}, and
\FpopN{}{\_}\
on stacks, and let $\FpopN{n}{\Qstates}$ stand for $n$ consecutive applications of
the function ${\rm{pop}}$ to the stack $\Qstates$.

\begin{definition}[\SRauta: execution \& language]
  Let $\mathcal{S}$ be an \SRauton\ with transition function $\tauSR$,
  and initial state $\Pinit$.
  The \emph{language accepted by} $\mathcal{S}$ is given by
  \begin{center}
    $\Flang{\mathcal{S}} =
     \{ w \mid \exists\, \Qstates,\Qprod
          \mbox{ such that }
          \trConfig{\epsilon}{\stacked{\Pinit}}{\epsilon}{\stacked{}} \trArrow^*
          \trConfig{\epsilon}{\Qstates}{w\$}{\Qprod}
          \}$
  \end{center}
  where the \emph{execution relation} $\trArrow$
  is defined by the following rules
  \begin{center}
  $\begin{array}{c}
    \prooftree
        \Ftop{\Qstates}=P , \quad
        \tauSR\pair{P}{x} = Q ,\quad
        h \in \set{\epsilon,\squared{x}}
    \justifies
        \trConfig{h}{\Qstates}{w}{\Qprod} \trArrow
        \trConfig{\epsilon}{\Fpush{Q}{\Qstates}}{wx}{\Qprod}
    \using
        (S)
    \endprooftree
  \\[4ex]
    \prooftree
        \Ftop{\Qstates}=P , \quad
        \tauSR\pair{P}{\trLabel{x}{R}{A\rightarrow\beta}} = Q  ,\quad
        \QstatesPrime = {\FpopN{\size{\beta}}{\Qstates}} , \quad
        \Ftop{\QstatesPrime} =R , \quad
        h \in \set{\epsilon,\squared{x}}
    \justifies
        \trConfig{h}{\Qstates}{w}{\Qprod} \trArrow
        \trConfig{\squared{x}}{\Fpush{Q}{\QstatesPrime}}{w}{\Fpush{A\rightarrow\beta}{\Qprod}}
    \using
        (R)
    \endprooftree
  \end{array}$
  \end{center}
\end{definition}

Rule (S) in the definition of $\trArrow$ mimics the execution of a shift move
of the shift-reduce algorithm.
Here we just observe two facts.
First, $h$ is required to be either empty or the prospective version
of the symbol which triggers the execution step.
Second, when $h$ actually equals $\squared{x}$, the application of rule (S)
obliterates the tracing of $\squared{x}$ in the reached configuration.
Hence, an execution step involving a prospective symbol $\squared{x}$ can never be followed
by a step depending on some $\tauSR\pair{P_n}{x'}$ with ${x'}\neq{x}$.
Also, the tracing of a prospective symbol stops once its concrete version
is met.
Rule~(R) simulates the execution of a reduce move, with the prospective symbol
$\squared{x}$ playing the lookahead $x$ that calls for the reduction.
Notice that the application of the (R) rule traces $\squared{x}$
in the first component of the new configuration.
Moreover, by the premiss $h \in \set{\epsilon,\squared{x}}$, an execution step
depending on the prospective symbol $\squared{x}$ can never be followed by
a step depending on $\squared{x'}$ if ${x'}\neq{x}$.
Overall, the intuition behind the use of prospective symbols in \SRauta\
is that once a symbol $\squared{x}$ starts playing as lookahead for a reduction,
it keeps doing so until an application of the (S) rule actually consumes
$\squared{x}$ in the first component of the configuration, and appends $x$
to the word under construction.

Next, we present a result on the correspondence between the execution of the
shift-reduce algorithm and the execution of \SRauta.

\begin{theorem}
\label{th:tableAndSR}
  Let $\mathcal{S}$ be the \SRauton\ for the parsing table
  $\pair{\mathcal{A}}{\Fla}$ of $\mathcal{G}$.
  Also, let $\Pinit$ be the initial state of $\mathcal{S}$.
  Then the following holds.
  \begin{itemize}
  \item
    If, for some local resolution of possible conflicts,
    the parsing of $w$ on $\pair{\mathcal{A}}{\Fla}$ is successful
    and returns $\Ftree{w}$ then,
    for some $\Qstates$,
    $\trConfig{\epsilon}{\stacked{\Pinit}}{\epsilon}{\stacked{}} \trArrow^*
     \trConfig{\epsilon}{\Qstates}{w\$}{\Ftree{w}}$.
  \item
    If, for some $\Qstates$ and some $\Qprod$,
    $\trConfig{\epsilon}{\stacked{\Pinit}}{\epsilon}{\stacked{}} \trArrow^*
     \trConfig{\epsilon}{\Qstates}{w\$}{\Qprod}$
    then, for some local resolution of possible conflicts,
    the parsing of $w$ on $\pair{\mathcal{A}}{\Fla}$ is successful
    and $\Ftree{w}=\Qprod$.
  \end{itemize}
\end{theorem}

\begin{pf}
  The first statement is proved by induction on the number of steps performed
  by the shift-reduce algorithm.
  The inductive handle is based on the following facts.
  Each move of the shift-reduce algorithm is matched by an execution step of
  the \SRauton, and the state-stack of the automaton evolves exactly as the parsing
  auxiliary stack.
  Also, at each move, the word-component of the configuration reached by the
  automaton equals the portion of input already processed by the shift-reduce
  algorithm, and the production-stacks of the two executions grow in lockstep
  fashion.

  In proving the second statement,
  care has to be taken to ensure that, if the \SRauton\ performs a step which involves
  either $x$ or $\squared{x}$, then the symbol $x$ is actually what the
  shift-reduce algorithm gets from its input buffer.
  To get the required guarantees, the statement is proven by induction on the
  length of \SRauta\ executions whose latest step is obtained by an application
  of rule~(S), which
  is the case, indeed, for executions leading to
  $\trConfig{\epsilon}{\Qstates}{w\$}{\Qprod}$.
  By the definition of $\trArrow$,
  if the execution step is inferred by rule~(S) and depends
  on an $x$-transition of the automaton, then $x$ shows as suffix of the
  word-component of the reached configuration.
  Suppose now that
  the execution step is deduced by rule~(R),
  that it involves $\squared{x}$, and
  that the word-component
  of the current configuration is $w$.
  If this is the case, then by the properties of $\trArrow^*$ described above,
  $x$ is the symbol that is appended to $w$ at the first (S)-step along the execution.
  As for the rest, the inductive handle is dual to that used in the proof
  of the first statement.
  It keeps track of the correspondence between pairs of stacks,
  and between the word-component of configurations and the word
  consumed by the shift-reduce algorithm.
\end{pf}

Cor.~\ref{cor:tableAndSR} below is an immediate consequence of
Th.~\ref{th:tableAndSR}.

\begin{corollary}
\label{cor:tableAndSR}
  Let $\mathcal{S}$ be the \SRauton\ for the parsing table
  $\pair{\mathcal{A}}{\Fla}$ of $\mathcal{G}$.
  Then $\Flang{\mathcal{S}}=\Flang{\mathcal{G}}$.
\end{corollary}

We conclude this section by a comment on nondeterminism.
As the proof of Th.~\ref{th:tableAndSR} hints, the nondeterminism of parsing
tables is reflected in \SRauta\ in a precise sense.
Suppose that the state $P$ contains a conflict on $x$.
If, during an execution of the \SRauton, $P$ is on top of the state-stack
when the current word is $w$, then there are two distinct ways
to reach a configuration for the word $wx$ by going along the distinct branches
outgoing $P$.

\section{Ambiguity detection}
\label{sec:detection}

In this section, we will present the proposed strategy for ambiguity detection.

The activity is carried on alternating two sorts of phases:
guessing and validation.
In the guessing phase, relying upon an approximation of \SRauta,
we identify words -- if any -- that might show the ambiguity of the grammar.
In the subsequent phase, we validate the paths associated with these words
against proper executions of \SRauta.

A grammar $\mathcal{G}$ is ambiguous iff there exists a word in $\Flang{\mathcal{G}}$
that has two distinct rightmost derivations, or, equivalently,
two distinct leftmost derivations.
The next result, that underpins the proposed detection strategy,
characterizes ambiguity in terms of executions of \SRauta.

Below, given any configuration $\Nconfig=\trConfig{h}{\Qstates}{w}{\Qprod}$,
we say that two execution steps from $\Nconfig$
\emph{are in conflict on $x$} to mean that
$P=\Ftop{\Qstates}$ has a conflict on $x$ and one of the following scenarios apply:
\emph{(i)} both the execution steps are inferred by rule~(R) and
  depend, respectively, on some
  $\tauSR\pair{P}{\trLabel{x}{R'}{p'}}$ and on some
  $\tauSR\pair{P}{\trLabel{x}{R''}{p''}}$
  such that $p' \neq {p''}$;
\emph{(ii)} one of the execution steps is inferred by rule~(S) and depends on
  $\tauSR\pair{P}{x}$, and the other is inferred by rule~(R) and depends on
  some $\tauSR\pair{P}{\trLabel{x}{R}{p}}$.

\begin{theorem}
\label{th:ambiguity}
  Let $\mathcal{S}$ be the \SRauton\ for the parsing table
  $\pair{\mathcal{A}}{\Fla}$ of $\mathcal{G}$.
  Also, let $\Pinit$ be the initial state of $\mathcal{S}$, and $\tauSR$
  be its transition function.
  Then $\mathcal{G}$ is ambiguous iff
  $w$ exists such that
  $\trConfig{\epsilon}{\stacked{\Pinit}}{\epsilon}{\stacked{}} \trArrow^* 
   \Nconfig$
  and
  $ \Nconfig \trArrow 
    \NconfigPrime \trArrow^* 
    \trConfig{\epsilon}{\Qstates_{1}}{w\$}{\Qprod_{1}}$ 
  and
  $  \Nconfig \trArrow 
    \NconfigDouble \trArrow^* 
    \trConfig{\epsilon}{\Qstates_{2}}{w\$}{\Qprod_{2}}$ 
  for some
  $\Qstates_{1},\Qprod_{1}$,
  $\Qstates_{2},\Qprod_{2}$,
  $\Nconfig,\NconfigPrime,\NconfigDouble$
  such that
  $\Nconfig \trArrow \NconfigPrime$ and $\Nconfig \trArrow \NconfigPrime$
  are in conflict on $x$.
\end{theorem}

\begin{pf}
(If)
If $\mathcal{G}$ is ambiguous then there exists $w\in\Flang{\mathcal{G}}$
that has two distinct rightmost derivations.
Hence by Th.~\ref{th:tableAndSR},
letting $\Pinit$ be the initial state of $\mathcal{S}$,
$\trConfig{\epsilon}{\stacked{\Pinit}}{\epsilon}{\stacked{}} \trArrow^* 
 \trConfig{\epsilon}{\Qstates_{1}}{w\$}{\Qprod_{1}}$
and
$\trConfig{\epsilon}{\stacked{\Pinit}}{\epsilon}{\stacked{}} \trArrow^* 
 \trConfig{\epsilon}{\Qstates_{2}}{w\$}{\Qprod_{2}}$
for some $\Qstates_{1},\Qstates_{2},\Qprod_{1},\Qprod_{2}$
with
$\Qprod_{1}\neq\Qprod_{2}$.
By $\Qprod_{1}\neq\Qprod_{2}$, the two executions above must differ at least for
one step inferred by the $(R)$-rule.
Since both executions lead to configurations with the same word-component,
the steps of the two executions are inferred from transitions of the
\SRauton\ involving the same sequence of symbols in $T\cup\set{\$}$.
Hence, the two executions must be made of zero or more equal steps up to
a configuration with a state $P$ at the top of the
state-stack that contains a conflict on some $x$.
From that configuration, the two executions must continue with distinct
steps that are in conflict on $x$.
\\
(Only if)
By Cor.~\ref{cor:tableAndSR}, $w\in\Flang{\mathcal{G}}$, and
by Th.~\ref{th:tableAndSR} both $\Qprod_{1}$ and $\Qprod_{2}$ represent
derivation trees for $w$.
It remains to show that $\Qprod_{1} \neq \Qprod_{2}$.
Assume $\Nconfig=\trConfig{h}{\Qstates}{w_1}{\Qprod}$.
If both $\Nconfig \trArrow \NconfigPrime$ and $\Nconfig \trArrow \NconfigDouble$
are inferred by the (R)~rule, then the production-stacks of
$\NconfigPrime$ and of $\NconfigDouble$ are obtained by pushing distinct
productions on $\Qprod$.
Nothing is popped out of production-stacks during execution,
hence the thesis.
Suppose now that $\Nconfig \trArrow \NconfigPrime$ is inferred by the (S)~rule,
and $\Nconfig \trArrow \NconfigDouble$ by the (R)~rule, and let
$A\rightarrow\beta$ be the production involved in the inference of
$\Nconfig \trArrow \NconfigDouble$.
Then, $w$ has the form $w'_{1}w''_{1}$ for some $w''_{1}$ that
is the frontier of the subtree rooted at $A$ in the tree described by $\Qprod_2$.
This is not the case for $w''_{1}$ in the tree represented by $\Qprod_1$.
Hence $\Qprod_{1} \neq \Qprod_{2}$.
\end{pf}

The labelled transition relations defined below are used to get approximations
of the language recognized by \SRauta.

\begin{definition}
  Let $\mathcal{S}$ be the \SRauton\ for the parsing table
  $\pair{\mathcal{A}}{\Fla}$ of $\mathcal{G}$.
  Also, let $\tauSR$ be the transition function of $\mathcal{S}$.
  For every pair of states $P$ and $Q$ of $\mathcal{S}$,
  $P \sArrow{x} Q$ iff $\tauSR\pair{P}{x} = Q$, and
  $P \sArrow{\squared{x}} Q$ iff $\tauSR\pair{P}{\trLabel{x}{R}{p}} = Q$
  for some $R$ and some $p$.
  Also,
  $\dArrow{x}$ stands for $\sArrowSqStar{\squared{x}} \sArrow{x}$, and
  $\dArrowWord{w}$ stands for $\dArrow{x_1}\ldots\dArrow{x_j}$
  for $w=x_1\ldots{x_j}$.
\end{definition}

To detect ambiguity, we analyze one conflict at a time, and look for words
that are accepted along paths taking either branch out of the conflict state.
The analysis is applied to all the conflicts of the \SRauton\ and searching
for longer and
longer words, up to the point that either we can conclude that the grammar is
ambiguous or unambiguous,
or a fixed bound on the length of the searched words is reached.
When analyzing a certain conflict of state $P$,
we assume that all the conflicts of the states other than $P$ are switched off
by applying a combination of resolutions for them.
If the analysis of the conflict of $P$ is inconclusive,
the analysis is retried for a different combination of resolutions for the
other conflicts.

The search for words that might have distinct derivations is carried on
alternating guessing and validation phases as described below.
In the guessing phase, we make use of the function $\Fguess{P}{Q}{l}$ that
returns the set of words $w$ shorter than $l$ and such that
$P\dArrowWord{w}Q$.
The validation phase checks whether an execution meeting specified
requirements exists.
If an execution from $\Nconfig$ to $\Nconfig'$ exists, where
$\Nconfig'$ is a configuration with $P$ at the top of its state-stack and with
$w$ as word-component, then the invocation of $\Fvalidate{\Nconfig}{w}{P}$
returns $\Nconfig'$, otherwise it returns failure.

For clarity, we first describe the main principles of the analysis of a
single conflict.
Then we comment on how inconclusive searches are handled.
We assume that the \SRauton\ at hand
has initial state $\Pinit$, and
final state $\Pfinal$.
Also, we let
$\Nconfig_0=\trConfig{\epsilon}{\stacked{\Pinit}}{\epsilon}{\stacked{}}$,
and let $l_1,l_2$ be integers.
We first consider the case that the state $P$ has an r/r conflict for $x$.
By Th.~\ref{th:ambiguity}, we focus on pairs of paths with the following shape
\begin{equation}
\label{eq:rr}
   \Pinit
    \dArrowWord{w_1} R
    \sArrowSqStar{\squared{x}}
    P
    \sArrow{\squared{x}} Z_1
    \dArrow{x} Q_1
    \dArrowWord{w_2\$} \Pfinal
    \quad
   \Pinit
    \dArrowWord{w_1} R
    \sArrowSqStar{\squared{x}}
    P
    \sArrow{\squared{x}} Z_2
    \dArrow{x} Q_2
    \dArrowWord{w'_2\$} \Pfinal
\end{equation}
where $Z_1$ and $Z_2$ are inferred by transitions of the automaton involving
$\squared{x}$ and two distinct productions $p_1$ and $p_2$.
We notice here that there are as many plausible instances of
$Z_i$ meeting the above requirements as the size of $\Fopening{p_i}{P}$.
To save on validation failures, we operate as follows.
\begin{enumerate}
\item
    For each $R$ such that $R\sArrowSqStar{\squared{x}}P$,
    we invoke $\Fguess{\Pinit}{R}{l_1}$.
    So, we can collect a set of triples
    $\triple{\Pinit}{w_1}{R}$ such that
    $\Pinit\dArrowWord{w_1}R$.
    Call \NsetOne\ such set.
\item
    For each triple $\triple{\Pinit}{w_1}{R}$ in \NsetOne, we invoke
    $\Fvalidate{\Nconfig_0}{w_1}{R}$.
    Call \NsetTwo\ the set of configurations obtained in this way.
    We use \NsetTwo\ to decide which are the most appropriate targets
    to consider among the $\squared{x}$-transitions outgoing $P$
    in (\ref{eq:rr}).
    To do that, we select the configurations in \NsetTwo\ that can undergo
    the following manipulation.
    We attempt to prolong the execution from each configuration in \NsetTwo\
    by performing zero or more steps driven by the (R)~rule under $\squared{x}$
    so to reach a configuration with $P$ on top of the state-stack.
    Then, by executing steps driven by the (R)~rule
    for $p_1$ and, resp., by the (R)~rule for $p_2$, we get configurations
    with $Z_i$ on top of the state-stack, with $i=1,2$.
    Next, we extend these executions further by means of zero or more steps driven by
    the (R)~rule under $\squared{x}$, and then by a step driven by the (S)~rule for $x$.
    So, from those configurations in \NsetTwo\ which can be extended as described
    above, we obtain pairs of configurations reachable from $\Nconfig_0$
    that have $w_1{x}$ as word-component and $Q_i$ at the top of their state-stacks.
    Call \NsetThree\ the set of these pairs.
\item
    For each pair $\pair{\Nconfig_1}{\Nconfig_2}$ in \NsetThree\
    we do the following.
    Suppose $Q_i$ is on top of the state-stack of $\Nconfig_i$.
    For each $w'\in\Fguess{Q_1}{\Pfinal}{l_2}$, we check whether
    $Q_2\dArrowWord{w'}\Pfinal$.
    If so, we run both
    $\Fvalidate{\Nconfig_1}{w'}{\Pfinal}$ and
    $\Fvalidate{\Nconfig_2}{w'}{\Pfinal}$.
    If both validations are successful, then the returned configurations
    have word-components $w_1{x}w'$ with $w'=w_2\$$ for some $w_2$.
    Also, their production-stacks
    represent two distinct derivation trees for $w_1{x}w_2$.
    Hence the grammar is ambiguous.
\end{enumerate}

Above, we streamlined the search strategy.
It remains to comment on the scenarios that induce us either to retry the analysis with
longer words, or to give up, or to conclude that the grammar is unambiguous.
They are identified as follows.
If either $\NsetOne$ or $\NsetTwo$ or $\NsetThree$ is empty, and if $l_1$ can be
increased further, then we try again from the beginning searching longer guesses for $w_1$.
Analogously, if either $\Fguess{Q_1}{\Pfinal}{l_2}$ is empty,
or no $w' \in \Fguess{Q_1}{\Pfinal}{l_2}$ is such that $Q_2\dArrowWord{w'}\Pfinal$,
or no $w' \in \Fguess{Q_1}{\Pfinal}{l_2}$ can be validated from both
$\Nconfig_1$ and $\Nconfig_2$,
then we can reuse $\NsetThree$ and retry with longer guesses for $w_2$.
This is reasonable, however, only under some circumstances.
Indeed, if all the possible guesses for $w_1$ and for $w_2$ have been computed
(and hence all the possible instances of $R$, $Z_1$ and $Z_2$ have been considered),
and if either the analyzed conflict is the single conflict of the grammar,
or analogous circumstances apply to all the conflicts of the grammar,
then we conclude that the grammar is unambiguous.

In case the state $P$ has an s/r conflict for $x$,
the relevant pairs of paths have the following shape
\begin{equation}
\label{eq:sr}
   \Pinit
    \dArrowWord{w_1} R
    \sArrowSqStar{\squared{x}}
    P
    \sArrow{x} Q_1
    \dArrowWord{w_2\$} \Pfinal
    \qquad\qquad
   \Pinit
    \dArrowWord{w_1} R
    \sArrowSqStar{\squared{x}}
    P
    \sArrow{\squared{x}} Z
    \dArrow{x} Q_2
    \dArrowWord{w'_2\$} \Pfinal
\end{equation}
and the analysis is carried on analogously to the case of an r/r conflict.
Shortly, we compute guesses for $w_1$ from $\Pinit$ to $R$,
then we validate the guesses against proper executions from $\Pinit$ to $R$,
and get extensions to $Q_1$ and to $Q_2$.
Next, we compute guesses for $w_2$ from $Q_1$ to $\Pfinal$,
then check whether they could, at least approximately, be matched from $Q_2$.
If so, we run the validations for $w_2$ from configurations whose
state-stacks have $Q_1$ and $Q_2$ at their top.

We conclude the section by playing the proposed strategy for the s/r
conflict on $\mathop{+}$ in state $5$
of the \SRauton\ for $\mathcal{G}_1$ (Fig.~\ref{fig:G1}).
Using the naming in (\ref{eq:sr}), $R$ ranges over $\set{1,5}$.
For $l_1=4$, we get
\begin{center}
  $\NsetOne=\set{a,a\mathop{+}a}$
\end{center}
and, by validation,
\begin{center}
$\NsetTwo=\{
  \langle {\epsilon},
          {\stacked{0,1}},
          {a},
          {\stacked{}}
  \rangle,
  \langle {\epsilon},
          {\stacked{0,2,4,1}},
          {a\mathop{+}a},
          {\stacked{p_2}}
  \rangle
  \}$.
\end{center}
The manipulation of $\trConfig{\epsilon}{\stacked{0,1}}{a}{\stacked{}}$ fails,
and we are left with 
\begin{center}
$\NsetThree=\{
  (
  \trConfig{\epsilon}{\stacked{0,2,4,5,4}}{a\mathop{+}a\mathop{+}}{\stacked{p_2,p_2}},
  \trConfig{\epsilon}{\stacked{0,2,4}}{a\mathop{+}a\mathop{+}}{\stacked{p_2,p_2,p_1}}
  )
  \}$.
\end{center}
Next, we compute guesses for $l_2=l_1$ from state $4$ to state $3$.
We obtain $a\$$ and run its validations from the configurations paired in
$\NsetThree$.
By that, we get
\begin{center}
$  \langle  {\epsilon},
            {\stacked{0,2,3}},
            {a\mathop{+}a\mathop{+}a\$},
            {\stacked{p_2,p_2,p_2,p_1,p_1}}
            \rangle$,
and
$  \langle  {\epsilon},
            {\stacked{0,2,3}},
            {a\mathop{+}a\mathop{+}a\$},
            $ $
            {\stacked{p_2,p_2,p_1,p_2,p_1}}
            \rangle$
\end{center}
that show the ambiguity of $\mathcal{G}_1$.

\section{Conclusions}
\label{sec:conclusions}

Starting from LR parsing tables, we defined \SRauta,
and used them to describe a conflict-driven strategy
to detect grammar ambiguity.
Through prospective symbols, lookaheads are accommodated on the edges of
\SRauta.
This feature was crucial to mine words in the language as paths on labelled graphs.

The reported strategy showed to be a quite handy way of reasoning about the ambiguity
of small grammars of scholarly size.
The assessment of its effectiveness for large grammars,
as well as comparisons with other detection methods,
is subject to
further investigation.
Other directions for future work are relative to possible applications of
\SRauta\ in testing the adequacy of the heuristics used by
parser generators to handle nondeterministic grammars.


\end{document}